\documentclass[a4paper,11pt]{article}
             
\usepackage{graphicx}
\usepackage{caption}
\usepackage{amsmath}
\usepackage{mathtools}
\usepackage{subcaption}
\usepackage{multirow}
\usepackage{jheppub} 

\usepackage[T1]{fontenc} 

\usepackage[numbers,sort&compress]{natbib}

\title{Nucleon and dinucleon decays to leptonic final states in a left-right symmetric model with large extra dimensions}

\author{Sudhakantha Girmohanta}
\affiliation{\ C. N. Yang Institute for Theoretical Physics and 
Department of Physics and Astronomy, \\
Stony Brook University, Stony Brook, NY 11794, USA }

\emailAdd{sudhakantha.girmohanta@stonybrook.edu}

\abstract{We consider baryon-number-violating nucleon and dinucleon decays to leptonic final states in the context of a left-right symmetric (LRS) model with large extra dimensions. Specifically, we study (a) nucleon to trilepton decays with $\Delta B=-1$ and $\Delta L=-3$, and (b) dinucleon to dilepton decays with $\Delta B=-2$ and $\Delta L=-2$.  In the LRS model, $B-L$ is gauged and is spontaneously broken by a Higgs vacuum expectation value $v_R$, which characterizes the scale at which processes violating $B-L$ occur. We show that together with the lower bound on $v_R$ from experimental limits on $n$-$\bar n$ oscillations,  constraints from searches for other nucleon decay modes imply sufficient suppression of these nucleon to trilepton and dinucleon to dilepton decay modes in this model to agree with experimental bounds.}

\usepackage{bm}
\newcommand{\beq}{\begin{equation}}
\newcommand{\eeq}{\end{equation}}
\newcommand{\beqs}{\begin{eqnarray}}\newcommand{\eeqs}{\end{eqnarray}}

\newcommand{\gsim}{\mathrel{\raisebox{-
.6ex}{$\stackrel{\textstyle>}{\sim}$}}}

\begin{document} 
\maketitle
\flushbottom


\section{Introduction}
\label{intro_section}

The Standard Model (SM), as modified to include neutrino masses and lepton mixing, is able to fit current experimental data. However, this model leaves many interesting questions unanswered. Baryon number ($B$) is a global symmetry of the Standard Model. Although $B$ is violated by non-perturbative effects in the SM \cite{hooft}, the rate of this violation is exponentially suppressed at temperature small compared with the electroweak scale. Furthermore, baryon-number violation (BNV) is one of the necessary conditions for explaining the observed baryon asymmetry in the Universe \cite{sakharov}, and it indeed occurs in many ultraviolet (UV) extensions of the standard model such as Grand Unified Theories (GUTs \cite{Georgi-Glashow,Minkowski_GUT, langacker_gut,NathPerez,Babu_BNV}).

One of the manifestations of BNV is nucleon decay, and this has been searched for since the early 1980s by many dedicated experiments. Here, we use the term ``nucleon decay'' to refer to baryon-number-violating decays of individual protons and bound neutrons in a nucleus having $\Delta B = -1$. Null results from these experiments have set stringent lower limits on the partial lifetimes of several nucleon decay modes \cite{pdg}. In the framework of low-energy effective field theory (EFT), nucleon decays have been analyzed by formulating SM-invariant four-fermion effective operators \cite{weinberg79,weinberg80,wz79,weldon-zee,abbott_wise} that connect three quarks with one lepton, thus mediating nucleon decays. Using the SM and a minimal framework of beyond-Standard Model (BSM) physics encoded in these four-fermion operators, ref. \cite{ndl} obtained improved lower bounds on the partial lifetimes of several nucleon decay modes.  

Neutron-antineutron ($n$-$\bar n$) oscillations have received substantial interest in the literature as a different class of BNV (\cite{mm80,chang,kuo_love,cowsik_nussinov,nnb82,nnb84}). These processes violate $B$ and $B-L$ (where $L$ denotes total lepton number) by two units and might provide the source for the observed baryon asymmetry in the universe \cite{kuzmin}. The effective Lagrangian for $n$-$\bar n$ oscillations, ${\cal L}^{(n \bar n)}_{eff}$, is non-diagonal in the $(|n\rangle, \ |\bar n \rangle)$ basis, reflecting the fact that the physical neutron state $|n\rangle_{phys.}$ is an admixture of both $|n\rangle$ and $|\bar n \rangle$ state. The $n-\bar n$ oscillations result from a nonzero transition amplitude, $\delta m \equiv |\langle \bar n | {\cal L}^{(n \bar n)}_{eff}| n \rangle|$. This has the consequence that an initial $nn$ or $np$ dinucleon state has a nonzero $\Delta B=-2$ decay into multi-meson final states, resulting in matter instability. This matter instability has been searched for in several experiments \cite{takita86,frejus,soudan,Akamland_nn_to_nunubar,sk_dinucleon_to_kaons,sk_nnbar,sk_dinucleon_to_pions,sno_nnb,bryman,takhistov15,sussman18,sno_invisible}. There has also been a search for $n$-$\bar n$ oscillations using a free neutron beam from a reactor \cite{ill}. Ref. \cite{dnd} used existing upper bounds on the decay rates for dinucleon decays to mesons to derive approximate upper bounds on the rates for $\Delta B=-2$, $\Delta L=0$ decays of dinucleon to dilepton final states (see also \cite{edn}). The estimated bounds obtained in \cite{dnd} are much more stringent than the current bounds from direct experimental searches
\cite{pdg,mcgrew99,skn}.

In contrast to nucleon decays, which are mediated by four-fermion operators, $n$-$\bar n$ oscillations are mediated by six-fermion operators. Hence, in four-space-time dimensions, the effective Lagrangian for nucleon decay, ${\cal L}^{(Nd)}_{eff}$, has coefficients of mass dimension $-2$, whereas ${\cal L}^{(n \bar n)}_{eff}$ has coefficients of mass dimension $-5$. Thus, naively, one would expect that $n$-$\bar n$ oscillations would be highly suppressed in comparison with single nucleon decay. Here one has to be careful in assuming a single mass scale responsible for BNV processes. The interplay of different mass scales may render operators having higher (free-field) mass dimension less suppressed than the operators having lower mass dimension. Ref. \cite{nnb02} provided an explicit example of this possibility. It considered an extra-dimensional framework \cite{as,ms} where fermions have strong localization in the extra-dimension(s). Operators are integrated over the extra-dimensions to obtain the long-distance effective Lagrangian in four-spacetime dimensions. This yields suppression factors in the effective Lagrangian originating as a result of the wave function separations of the fields involved in the higher dimensions.  The authors in ref. \cite{nnb02} demonstrated that although separating the wave functions of quarks and leptons in the higher dimension(s) suppresses the rate for single nucleon decay far below current observable resolution, this does not suppress $n$-$\bar n$ oscillations, which only require the overlap among quark wave functions. Thus, in this case $n$-$\bar n$ oscillations and the resulting dinucleon decays are the main manifestation of baryon number violation.

A more general question was asked in ref. \cite{bvd}, namely in an extra-dimensional framework, can operators having $n_{f_2}$-fold fermions be less suppressed than operators comprising of $n_{f_1}$-fold fermions even if $n_{f_2} > n_{f_1}$? In other words, given the constraint of wave function separation of quarks and leptons in the higher dimensions from single nucleon decay, are other nucleon and dinucleon decay modes suppressed enough so that they agree with the current experimental bounds? We explored this question in a Standard Model effective field theory (SMEFT) framework with large extra dimensions and considered various nucleon and dinucleon decay modes \cite{bvd}.

In this paper we investigate the above-mentioned question in the context of left-right symmetric (LRS) gauge group:
\beq
G_{LRS} \equiv SU(3)_c \otimes SU(2)_L \otimes SU(2)_R \otimes U(1)_{B-L} \ .
\label{GLRS_eqn}
\eeq
We focus on (a) nucleon decays to trilepton final states with $\Delta B=-1$ and $\Delta L=-3$, and (b) dinucleon decays to dilepton final states with $\Delta B=-2$ and $\Delta L=0$. 
Our present work complements our previous study in \cite{nnblrs}, where we showed that, in contrast to the analysis in \cite{bvd}, which was performed in a SMEFT framework, some of the operator mediating $n$-$\bar n$ oscillations in the LRS framework yield no suppression from the extra-dimensional integration, thus rendering them even less suppressed. This motivates further experimental searches for $n$-$\bar n$ oscillations and resultant matter instability. Indeed, these searches will be carried out at DUNE (Deep Underground Neutrino Experiment) 
\cite{DUNE_abi,dune_physics,barrow} and Hyper-Kamiokande \cite{Abe_HK}, and there are also prospects for an 
$n-\bar n$ search experiment at the European Spallation Source \cite{nnbar_physrep}. 

As the field content and their representations under $G_{LRS}$ are different than those in the SMEFT, it is useful to consider various nucleon and dinucleon decay modes in the theory and check if they are consistent with current experimental data. We carry out this task here. In the framework of effective field theory, we formulate different operator classes mediating various BNV nucleon and dinucleon decay modes and estimate the resultant decay rates. It is also of interest to consider color-nonsinglet scalar fields mediating BNV \cite{nnb84,wise}, but here in our low-energy effective Lagrangian approach, we restrict ourselves to the minimal LRS fermion and Higgs field content defined in table \ref{LRS_field_content_table}. Although we work in the context of a model with large extra dimensions here, our operators have corresponding operators in four-spacetime dimensions, the difference being that the fermions would have no additional higher-dimensional profile in that case. Hence our study is relevant for applying the $LRS$ model to analyze the physics of baryon number violation.

The organization of this paper is as follows: in section \ref{theory_section} we explain the theoretical framework that we use. In section \ref{nd3_section} we discuss various decay modes of nucleons to trilepton final states. In section \ref{dinucleon_section} various dinucleon decays to dileptons are considered. We conclude in section \ref{conclusions_section}. In the appendices \ref{intform_appendix}, \ref{decay_width_appendix}, \ref{operators_Nm3_appendix} we include some useful formulas and some further details of our calculations. 

\section{Theoretical framework}
\label{theory_section}

We work in the LRS model with large extra dimensions, which is similar to the model used in ref. \cite{nnblrs}. To be self-contained, we will recapitulate this briefly and direct the reader to refs. \cite{nnblrs,bvd} and references therein for more details. 

\subsection{LRS model}\label{LRS_section}

Let us recall some salient features of the LRS model \cite{mm80,lrs75a,lrs75b,lrs81}. The Lagrangian is invariant under the gauge group $G_{LRS}$ defined in eq. \eqref{GLRS_eqn}. The fermion and Higgs fields in the theory and their representation under $G_{LRS}$ gauge group are shown in table \ref{LRS_field_content_table}. 
%
\begingroup
\setlength{\tabcolsep}{10pt} 
\renewcommand{\arraystretch}{1.2} 
\begin{table}[ht]
\begin{center}
\begin{tabular}{|c| c c c|} \hline\hline
Fields &Notation & Explicit form &  $G_{LRS}$ representation\\ 
\hline
\multirow{8}{4em}{Fermions}  &$Q^{i \, \alpha}_{L,g}$     & $\begin{pmatrix} u^{\alpha} \\ d^{\alpha}\end{pmatrix}_{L,g}$ & $(3,2,1)_{\frac{1}{3}}$\\[10pt]
                            &                            $Q^{i^\prime \, \alpha}_{R,g}$ & $\begin{pmatrix} u^{\alpha} \\         d^{\alpha}\end{pmatrix}_{R,g}$& $(3,1,2)_{\frac{1}{3}}$\\[15pt]
                            &$L^i_{L,g}$         & $\begin{pmatrix} \nu \\ \ell\end{pmatrix}_{L,g}$     & $(1,2,1)_{-1}$\\[15pt]
                            & $L^{i^\prime}_{R,g}$   & $\begin{pmatrix} \nu \\ \ell\end{pmatrix}_{R,g}$      & $(1,1,2)_{-1}$\\[15pt]
                            \hline 
\multirow{6}{4em}{Higgs}    & $\Phi^{i j^\prime}$                      & $\begin{pmatrix}
                                                                        \phi_1^{0} & \phi_1^{+} \\
                                                                         \phi_2^{-} & \phi_2^{0}
                                                                                        \end{pmatrix}$ & $(1,2,2)_{0}$\\[20pt]
                             & $\Delta^{i j}_L$                     & $\begin{pmatrix}
                                                                         \Delta_L^{+}/\sqrt{2} & \Delta_L^{++} \\
                                                                        \Delta_L^{0} & -\Delta_L^{+}/\sqrt{2}
                                                                        \end{pmatrix}$                   & $(1,3,1)_{2}$\\[20pt]
                            & $\Delta^{i^\prime j^\prime}_R$              & $\begin{pmatrix}
                                                                        \Delta_R^+/\sqrt{2} & \Delta_R^{++} \\
                                                                         \Delta_R^0 & -\Delta_R^+/\sqrt{2}
                                                                        \end{pmatrix}$                  & $(1,1,3)_{2}$\\[20pt]
\hline\hline
\end{tabular}
\end{center}
  \caption{Fermion and Higgs fields in the LRS model and their representations under the $G_{LRS}$ group.}
\label{LRS_field_content_table}
\end{table}
\endgroup
Here, $SU(3)_c$ indices are indicated by Greek letters $\alpha, \beta ...$; Roman unprimed letters, such as $i,j ...$, are $SU(2)_L$ indices, and primed Roman letters $i^\prime,j^\prime ...$ denote $SU(2)_R$ indices. The subscript $g$ on fermion fields is the generation index, i.e., $u_g = u,c,t$ and $\nu_g = \nu_e, \nu_\mu, \nu_\tau$ for $g=1,2,3$ respectively, and similarly for $d_g$, and $\ell_g$. In the fourth column of table \ref{LRS_field_content_table}, the three numbers in each parenthesis are the dimensionalities of the field representation with respect to the three non-Abelian factors in $G_{LRS}$, and the subscript is the $B-L$ charge. The subscripts $L,R$ in a fermion field denote the chirality of the fermion (and this $L$ should not be confused with the lepton number). The electromagnetic charge $Q_{em}$ is expressed by this elegant formula
\beq \label{qem_lrs}
    Q_{em} = T_{3L} + T_{3R} + \frac{B-L}{2} \ ,
\eeq
 where ${\vec{T}}_{L,R}$ are the $SU(2)_{L,R}$ generators.
 
 Several studies have investigated the minimization of the Higgs potential \cite{dmrx,dgko,lrs81}. Spontaneous symmetry breaking of the $G_{LRS}$ group takes place sequentially as various Higgs fields pick up vacuum expectation values (VEVs).  Since no right-handed weak charged current processes have been 
 observed experimentally, and rather stringent upper limits have been set on them, the highest scale of gauge symmetry breaking occurs with $\Delta_R$ picking a VEV $\langle \Delta_R\rangle_0 = v_R$, thereby breaking $G_{LRS} \to G_{SM}$.  At the electroweak symmetry-breaking (EWSB) scale of $\simeq 250$ GeV, $\Phi$ picks up VEVs, breaking $G_{SM} \to {\rm SU}(3)_c \otimes {\rm U}(1)_{em}$.  The VEV of $\Delta_L$, $v_L$, is much smaller than the EWSB scale, as constrained by the closeness of the parameter $\rho = m_W^2/(m_Z^2\cos^2\theta_W$) to unity.  After appropriate rephasings, the VEVs of the Higgs fields can be written down in the following form:
    %
    \begin{align}
        \langle \Phi \rangle_0 &= \begin{pmatrix}
              \kappa_1 &  0 \\
               0       & \kappa_2 e^{i\theta_\Phi}
                                   \end{pmatrix} \ ,  
       &\langle \Delta_L \rangle_0 &= \frac{1}{\sqrt{2}} \begin{pmatrix}
                0       &  0 \\
                v_L e^{i\theta_{\Delta}} & 0
                                    \end{pmatrix} \ ,
        & \langle \Delta_R \rangle_0 &= \frac{1}{\sqrt{2}} \begin{pmatrix}
                 0   &  0 \\
                v_R &  0 
                                    \end{pmatrix}\ .
        \label{VEV_LRS_eqn}
    \end{align}
As mentioned above, experimental results imply the inequalities $v_L \ll \kappa_1,\kappa_2 \ll v_R$. Supersymmetric extensions of the LRS model are also of interest (e.g., \cite{susylrs}), but as the LHC has not yet observed any supersymmetric particles, we restrict ourselves to the non-supersymmetric version in this analysis.

\subsection{Extra-dimensional framework}\label{dimension_section}

An interesting feature of the extra-dimensional model of ref. \cite{as} is that hierarchies in the long-distance theory are not generated by symmetries in the short-distance theory, but rather originate from the ``geography'' of fermion wave functions in the extra dimensions. Standard-Model fields are considered in $d=4+n$ spacetime dimensions, where the $n$ extra (spatial) dimensions are chosen to be compact. The lowest Kaluza-Klein mode (zero-mode) of a given SM fermion has strong localization at a certain point in the extra dimensions. In particular, fermion wave functions have Gaussian profiles in the extra dimensions. This localization can be achieved by coupling the fermions in an appropriate way with scalars \cite{rubakov83,kaplan_schmaltz,dvali_shifman,surujon}. This framework has the attractive feature that it can explain the very large range of SM fermions with ${\cal O}(1)$ Yukawa coupling in the higher dimensional theory. The way that the fermion mass hierarchy arises in this type of model is via differences in the distances between centers of left-chiral and right-chiral components of the fermion wave functions in the higher dimensions. 

We use a Wilsonian effective field theory (EFT) framework. Because the length scale of the extra dimensions is very small, one integrates over the extra dimensions to obtain the long-distance four-dimensional effective Lagrangian. Here, $x_\mu$ ($\mu = 0,1,2,3$) denotes the usual 4-spacetime dimensions, whereas $y_\nu$ ($\nu = 1,2,...,n$) represents coordinates in the extra-dimensions. The fields are restricted to an interval $0 \leq y_\nu \leq L$ in the extra dimensions. The extra dimensions extend over a length scale $L$, and thus the corresponding energy scale is $\Lambda_L \equiv L^{-1}$. The wave function for a general fermion $f$ is taken to have the factorized form 
\beq
\Psi_f (x,y) = \psi_f (x) \ \chi_f (y) \ .
\label{fermion_extra_dimension_eqn}
\eeq
Here $\chi_f (y)$, the fermion wave function in the extra dimensions, is Gaussian in shape \cite{as,ms}:
\beq
\chi_f (y) = A \ e^{-\mu^2 \|y-y_f\|^2} \ ,
\label{gaussian_eqn}
\eeq
where $y_f \in {\mathbb R}^n$ specifies the position vector of the fermion $f$ in the extra-dimensional space, with components $y_f = (y_{f_1}, y_{f_2},...,y_{f_n})$, and $ \| y_f \| \equiv (\sum_{i=1}^n y_{f_i}^2)^{1/2}$ is the usual Euclidean norm of the vector $y$. The fermion has strong localization in the sense that $L_\mu \equiv \mu^{-1} \ll L$. Thus while integrating over the extra dimensions to obtain the low energy effective Lagrangian involving fermion fields, we can safely approximate the interval of integration as $(-\infty,\infty)$ instead of $[0,L]$. In order to reproduce the canonical normalization for the kinetic term for the fermion in the low energy effective Lagrangian, the prefactor $A$ in eq. \eqref{gaussian_eqn} is 
\beq
  A = \bigg ( \frac{2}{\pi} \bigg )^{n/4}\, \mu^{n/2} \ . 
  \label{A_eqn}
\eeq
Constraints from flavor-changing neutral-current processes \cite{pdg}, precision electroweak data, and collider searches \cite{dpq2000,barenboim_fcnc,acd,abpy} can be accommodated by making $\Lambda_L \gsim 100$ TeV. The ratio $L/L_{\mu} = \mu/\Lambda_L \sim 30$ is chosen to provide enough space for separation of fermion wave functions in the extra-dimensions.  Thus, we take $\mu \sim 3 \times 10^3$ TeV. Let us define a dimensionless wave function separation parameter
\beq
\eta = \mu \, y \ .
\label{eta_definition_eqn}
\eeq

Higgs fields are assumed to have flat profiles in the extra dimensions \cite{as,surujon}. As there is only one scale for this flat profile, namely the size of the extra dimensions, $L$, the dependence on $L$ drops out in the four-dimensional effective Lagrangian for the scalar fields. This can be shown formally by the use of box-normalization methods \cite{nnblrs}. This is in contrast with the fermions, which have an intrinsic localization scale $\mu$ characterizing the width of their Gaussian wave function profiles. Accordingly, the scalar obtaining a VEV in an operator can simply be replaced by its VEV in four-spacetime dimensions.

Although $B-L$ is gauged in the LRS model, the VEV $v_R$ of the Higgs field $\Delta_R$ spontaneously breaks $B-L$ (by two units), and gives rise to various processes with $|\Delta(B-L)|=2$. Let us say that we are interested in a process `$\cal X$', which, at the lowest possible mass dimension, is mediated by operators ${\cal O}^{(\cal X)}_r (x,y)$ comprised of $n_f$ fermion fields and $n_{\Delta_R}$ right-handed triplet Higgs fields. The mass scale characterizing the physics responsible for this process is denoted by $M_{\cal X}$. We note that for the case for nonzero $n_{\Delta_R}$, which must be the case if the process violates $B-L$, then 
$M_{\cal X}=v_R$. Thus the corresponding Lagrangian in $d=4+n$ spacetime dimensions is 
\beq
{\cal L}^{(\cal X)}_{eff, 4+n} (x,y) = \sum_r \kappa_{r}^{(\cal X)} \ {O}^{(\cal X)}_r (x,y) \ .
\label{Leff_d_dim_eqn}
\eeq
Possible generational dependence of fermion fields has been suppressed in the notation. We avoid explicitly writing the dependence of the coefficients on $n_f$ and $n_{\Delta_R}$ to make the notation less cumbersome; rather, this dependence is understood implicitly by the presence of the superscript ${\cal X}$. Integrating over the extra dimensions, we get the effective Lagrangian in four-spacetime dimensions:
\beq
{\cal L}^{(\cal X)}_{eff} (x) = \sum_r c_{r}^{(\cal X)} \ {\cal O}^{(\cal X)}_r (x) \ .
\label{Leff_4dim_eqn}
\eeq
Using dimensional analysis, eqs. \eqref{A_eqn} and \eqref{intform} we can relate the two coefficients:
\beq \label{c_kappa_relation_eqn}
\begin{split}
c_{r}^{(\cal X)} &= \kappa_{r}^{(\cal X)} \ A^{n_f} \ \mu^{-n} \ \Big(\frac{\pi}{n_f}\Big)^{n/2}\ e^{-S_{r}^{(\cal X)}} \ (v_R)^{n_{\Delta_R}}  \\
                         &= \frac{{\bar \kappa}_{r}^{(\cal X)}}{M_{\cal X}^{(3n_f-8)/2}} \, \bigg( \frac{v_R}{M_{\cal X}} \bigg)^{n_{\Delta_R}} \, \bigg(\frac{\mu}{M_{\cal X}}\bigg)^{(n_f-2)n/2} \, \Bigg( \frac{2^{n_f/4}}{\pi^{(n_f-2)/4} \, n_f^{1/2}} \Bigg)^n \, e^{-S_{r}^{(\cal X)}}  \ . 
\end{split}
\eeq
In the second line we have extracted the mass dependence of the coefficient $\kappa_r^{(\cal X)}$ to define a new dimensionless ${\cal O}(1)$ coefficient ${\bar \kappa}_r^{(\cal X)}$. The exponential suppression coming from the integration is denoted as $e^{-S_r^{(\cal X)}}$. 
We will use the constraints on the wave function separation coming from the nucleon decay and quark masses and also the constraint on $v_R$ from $n$-$\bar n$ oscillations (eqs. (3.16) , (4.17) and (5.19) in ref. \cite{nnblrs}) to determine the degree of suppression of various baryon-number-violating decay modes.  
%
%
\section{Nucleon decays to trileptons with \boldmath $\Delta B=-1$ and \boldmath $\Delta L=-3$}
\label{nd3_section}

In this section we consider BNV processes having $\Delta B = -1$ and $\Delta L = -3$, such as $p \to \ell^+ \bar \nu \bar \nu^\prime$ and $n \to \bar \nu \bar \nu^\prime \bar \nu^{\prime \prime}$, where $\ell$ is a lepton which is allowed by the available phase space, i.e, $e$ or $\mu$. The prime in the $\nu^\prime$ represents the fact that it can be of a different generation than $\nu$. The operators mediating these decays contain three quark fields and three lepton fields, and hence are six-fermion operators.  Since these decays violate $B-L$ by 2 units, they can arise from the VEV of a $\Delta_R$ field, so minimal operators will involve a single $\Delta_R$ field.  Our analysis here is complementary to analyses using a SMEFT in four-dimensional spacetime such as \cite{heeck_takhistov}.

In passing, we note that one can, in principle, consider corresponding processes related by crossing.  For example, corresponding to the proton decay mode $p \to \ell^+ \bar \nu \bar \nu^\prime$, there is the crossed process 
$\nu p \to \ell^+ \bar \nu^\prime$, which also has $\Delta B=-1$ and $\Delta L=-3$. However, the rate for this would be negligibly small and there would be a much larger background from antineutrino contamination in a beam comprised mainly of neutrinos, in addition to which one would not observe the outgoing $\bar \nu^\prime$, so this would not provide a feasible way to search for this type of $B$ and $L$ violation. We shall refer to $\Delta B=-1$, $\Delta L=-3$ nucleon decay modes as
`$Nm3$', where the acronym stands for ``nucleon decay to trileptons having $\Delta L$ equal minus 3''.  As outgoing neutrinos are not observed in an experiment, in the context of LRS model, they can be either $\bar \nu_{L,g}$ or $\bar \nu_{R,g}$. We subsume all these possibilities by generically calling them as $\bar \nu$. 

Here it is worthwhile to point out a difference between this analysis and the analysis in ref. \cite{bvd}, where these decay modes were considered in a SM effective field theory framework. The mass scale $M_{Nm3}$ in SMEFT characterizing nucleon decay to trileptons is unknown, in the sense that it is not related to any parameters in the theory. On the other hand $B-L$ is gauged as $U(1)_{B-L}$ in the LRS model and is spontaneously broken by the nonzero VEV, $v_R$, of the Higgs field $\Delta_R$. Consequently, $Nm3$ processes, which violate $B-L$ by two units, have a characterizing mass scale $M_{Nm3} = v_R$ for the LRS model, which is the same mass scale that is relevant for $n$-$\bar n$ oscillations. Therefore, we can use the experimental bounds on $n$-$\bar n$ oscillations to constrain the rates of $Nm3$ processes.

As the exponential suppression factor, $e^{-S_r}$, is determined by the number of fermion fields present in an operator (see eq. \eqref{intform}), we focus on classes of operators characterized by having the same field content and hence the same suppression factor. A given class will generically contain operators that may contribute to different processes, but each class will have a particular suppression factor from the integration of fermion fields over the extra dimensions. We describe here how to get the general classes for $Nm3$ processes. 

A general class of operators is determined by the number of fields present. We look for the lowest-dimensional classes of operators, $C^{(Nm3)}_r$, that can mediate $Nm3$ processes. Suppressing the generational indices, an operator of this type, denoted $C^{(Nm3)}_r$, can be written down as follows:
\beq \label{cnm3_eqn}
C^{(Nm3)}_r \sim Q_L^{n_{Q_L}} \, Q_R^{n_{Q_R}} \, L_L^{n_{L_L}} \, L_R^{n_{L_R}} \, \Delta_R^{n_{\Delta_R}} \ .
\eeq
The notation $\psi^{n_\psi}$ means there are $n_\psi$ number of $\psi$ fields present in a class. An operator  $C^{(Nm3)}_r$ must be Lorentz-invariant, and, before the spontaneous breaking of the $G_{LRS}$ gauge symmetry, it should also abe invariant under $G_{LRS}$ gauge transformations, since it is envisioned as arising in a UV theory that is $G_{LRS}$-invariant. As we are interested in processes which involve three quarks and three leptons, we require:
\begin{align}
    n_{Q_L} + n_{Q_R} &= 3 \label{Nm3_nq_eqn} \\
    n_{L_L} + n_{L_R} &= 3 \label{Nm3_nl_eqn} \ .
\end{align}
Eq. \eqref{Nm3_nq_eqn} automatically satisfies the SU(3)$_c$ triality condition. In other words, with an appropriate contraction with the (totally antisymmetric) $\epsilon_{\alpha \beta \gamma}$ color tensor, 
a product of three quark fields can always be contracted to form an invariant with respect to color SU(3)$_c$.  A necessary and sufficient condition for invariance under $U(1)_{B-L}$ is 
\beq \label{Nm3_b-l_eqn}
    \frac{1}{3} (n_{Q_L} + n_{Q_R}) - (n_{L_L}+n_{L_R})+ 2 n_{\Delta_R} = 0 \ .
\eeq
Together with eqs. \eqref{Nm3_nq_eqn} and \eqref{Nm3_nl_eqn}, this implies that
\beq \label{Nm3_ndelta_eqn}
    n_{\Delta_R} = 1 \ .
\eeq
As we need an even number of left-handed fields to form an SU(2)$_L$ singlet, we need
\beq \label{Nm3_su2l_eqn}
   \widetilde{N}_L \equiv  n_{Q_L} + n_{L_L} = \textrm{even} \ .
\eeq
Since we have $n_{\Delta_R} = 1$, and as $\Delta_R$ transforms as an isotriplet under SU(2)$_R$, it 
follows that, in order for $C_r^{(Nm3)}$ to be SU(2)$_R$-invariant, we need to have the contraction of the other right-handed fields in the operator also form an isotriplet. Hence,
\beq \label{Nm3_su2r_eqn}
   \widetilde{N}_R \equiv n_{Q_R} + n_{L_R} = \textrm{non-zero even} \ .
\eeq
Adding eqs. \eqref{Nm3_nq_eqn}, \eqref{Nm3_nl_eqn}, one has $(n_{Q_L} + n_{L_L}) + (n_{Q_R}+n_{L_R})=6$, so the condition 
\eqref{Nm3_su2l_eqn} is implied by set of conditions in eqs. \eqref{Nm3_nq_eqn}, \eqref{Nm3_nl_eqn}, and \eqref{Nm3_su2r_eqn}. Since we are looking for the lowest-dimensional operator, we do not consider derivatives in the operators. Hence the requirement of Lorentz invariance implies 
\begin{align} 
    n_{Q_L} + n_{L_L} &= \textrm{even} \label{Nm3_lorentz_l_eqn} \\
     n_{Q_R} + n_{L_R} &= \textrm{even} \label{Nm3_lorentz_r_eqn} \ .
\end{align}
Clearly, eqs. \eqref{Nm3_su2l_eqn} and \eqref{Nm3_su2r_eqn} satisfy the Lorentz invariance condition for the case in which the operator does not contain any derivatives. 

We note that if there is an odd number of derivatives present in the operator, then the Lorentz invariance condition would demand that the right hand sides of eqs. \eqref{Nm3_lorentz_l_eqn}, \eqref{Nm3_lorentz_r_eqn} be changed to `odd', thus contradicting eqs. \eqref{Nm3_su2l_eqn}, \eqref{Nm3_su2r_eqn}. Consequently, there can be only even number of derivatives in this case.  Enumerating positive integer solutions of the linearly independent equations, namely eqs. \eqref{Nm3_nq_eqn}, \eqref{Nm3_nl_eqn}, \eqref{Nm3_ndelta_eqn}, and \eqref{Nm3_su2r_eqn}, gives us the classes mediating $Nm3$ processes. These are listed in Table \ref{Nm3_class_table}.
%
%
\begingroup
\setlength{\tabcolsep}{10pt} 
\renewcommand{\arraystretch}{1.5} 
    \begin{table}[ht]
        \begin{center}
            \begin{tabular}{|c|c|c|c|} \hline\hline
Class $C_{r}^{(Nm3)}$  & $\widetilde{N}_L$ & $\widetilde{N}_R$  & Structure   \\ \hline
        $C_1^{(Nm3)}$  & $0$ & $6$       & $Q_{R}^3 \,  L_R^3 \, \Delta_R$ \\
        $C_2^{(Nm3)}$  & $2$ & $4$       & $Q_R^3 \, L_L^2 \, L_R \, \Delta_R$ \\
        $C_3^{(Nm3)}$  & $2$ & $4$       & $Q_L \, Q_R^2 \, L_L \,L_R^2 \, \Delta_R$ \\
        $C_4^{(Nm3)}$  & $2$ & $4$       & $Q_L^2 \, Q_R \, L_R^3 \, \Delta_R$ \\
        $C_5^{(Nm3)}$  & $4$ & $2$       & $Q_L^2 \, Q_R \, L_L^2 \, L_R \, \Delta_R$ \\
        $C_6^{(Nm3)}$  & $4$ & $2$       & $Q_L \, Q_R^2 \, L_L^3 \, \Delta_R$ \\
        $C_7^{(Nm3)}$  & $4$ & $2$       & $Q_L^3\, L_L \, L_R^2 \, \Delta_R$ \\
                \hline \hline
            \end{tabular}
        \end{center}
            \caption{List of classes of operators mediating $Nm3$ processes. The numbers $\widetilde{N}_L$ and $\widetilde{N}_R$ denote the total numbers of left-handed and right-handed fermion doublets present in the class.}
    \label{Nm3_class_table}
\end{table}
\endgroup

\begin{figure}[ht]
    \begin{subfigure}{0.45\textwidth}
        \includegraphics[width=\hsize]{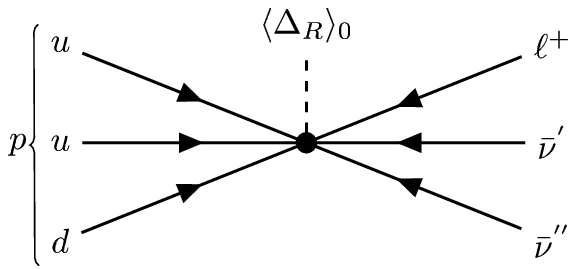}
        \caption{}
      \label{ptolnunu_figure}
    \end{subfigure}
    \vspace{1cm}
\hfill
    \begin{subfigure}{0.45\textwidth}
    \includegraphics[width=\hsize]{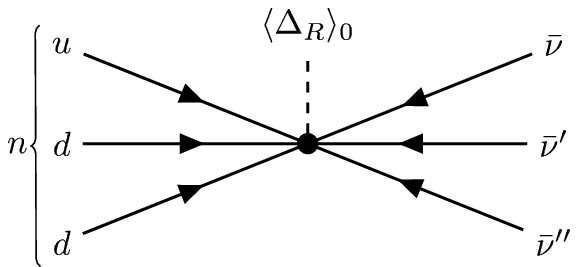}
    \caption{}
          \label{ntonununu_fig}
\end{subfigure}
\vspace{1cm}
    \begin{subfigure}{0.45\textwidth}
    \includegraphics[width=\hsize]{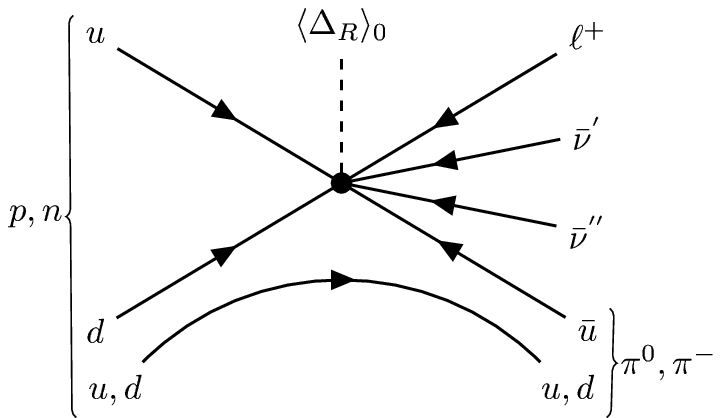}
    \caption{}
    \label{Ntopilnunu_fig}
    \end{subfigure}
    \hfill 
     \begin{subfigure}{0.45\textwidth}
    \includegraphics[width=\hsize]{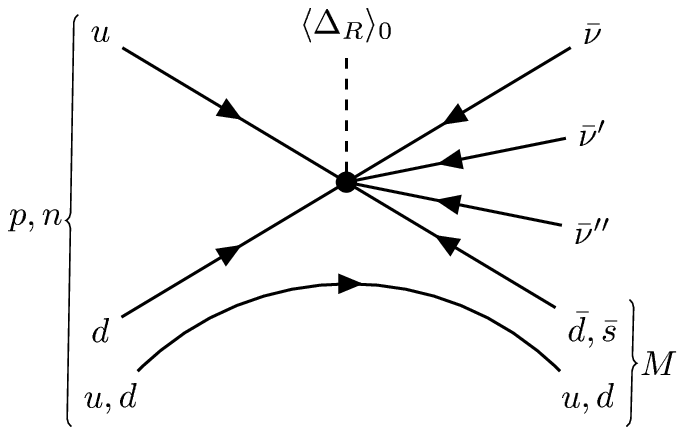}
    \caption{}
    \label{NtonununuM_fig}
    \end{subfigure}
    \caption{Some BNV processes mediated by ${\cal O}_{1,{C_1^{(Nm3)}}}$. Figures \ref{ptolnunu_figure}, \ref{ntonununu_fig} represent the three-body decay modes, whereas Figures \ref{Ntopilnunu_fig}, \ref{NtonununuM_fig}
    display some of the four-body decay modes that are produced by the same type of BNV operator. Here, $M$ means meson.}
    \label{Nm3_figure} 
\end{figure}

Each separate class contains several operators differing by group contraction tensors, which hence mediate different processes. Let us illustrate this by constructing some explicit operators. Consider an operator belonging to $C_1^{(Nm3)}$:
    \begin{multline}
    {\cal O}_{1,{C_1^{(Nm3)}}} =  \Big[ Q_{R,g_1}^{i^\prime \alpha T} \, C \, Q_{R,g_2}^{j^\prime \beta} \Big] \, \Big[ Q_{R,g_3}^{k^\prime \gamma T} \, C \, L_{R,g_4}^{m^\prime} \Big] \, \Big[ L_{R,g_5}^{p^\prime  T} \, C \, L_{R,g_6}^{q^\prime} \Big] (\Delta_R)^{r^\prime s^\prime}\\
                                 \times \epsilon_{\alpha \beta \gamma} \, \epsilon_{i^\prime j^\prime} \, \epsilon_{k^\prime m^\prime} \, (\epsilon_{p^\prime r^\prime} \, \epsilon_{q^\prime s^\prime} + \epsilon_{p^\prime s^\prime} \, \epsilon_{q^\prime r^\prime}) \ . 
                            \label{O1_nm3_eqn}
    \end{multline}
Here, $C$ is the Dirac charge conjugation matrix with the properties $C^T = -C$ and $C \gamma_\mu C^{-1} = -(\gamma_\mu)^T$. From eq. \eqref{qem_lrs}, it is clear that the neutral component of $\Delta_R$, which gets a VEV $v_R = \langle \Delta_R \rangle_0$, has $T_{3R} = -1$. Therefore, the lepton bi-doublet, which is contracted symmetrically with $\Delta_R$ to form $SU(2)_R$ singlet, must possess $T_{3R} = +1$ . Hence, writing eq. \eqref{O1_nm3_eqn} explicitly, we have 
    \begin{multline}
        {\cal O}_{1,{C_1^{(Nm3)}}} =  \epsilon_{\alpha \beta \gamma} \, \Big[ u_{R,g_1}^{\alpha T} C d_{R,g_2}^{\beta}- d_{R,g_1}^{\alpha T} C u_{R,g_2}^{\beta}\Big]  \Big[ u_{R,g_3}^{\gamma T} C \ell_{R,g_4}-d_{R,g_3}^{\gamma T} C \nu_{R,g_4}\Big]  \\
         \times 2 \, \Big[\nu_{R,g_5}^{T} C \nu_{\ell,R,g_6}\Big] \langle \Delta_R \rangle_0  \ .
        \label{explicit_o1_eqn}
    \end{multline}
Suppressing the color indices, fermion chiralities, neutrino generation indices, and charge conjugation matrix, schematically, the operator in eq. \eqref{O1_nm3_eqn} contains $u_{g_1} d_{g_2} u_{g_3} \ell_{g_4} \nu^2 \langle \Delta_R \rangle_0$ and $u_{g_1} d_{g_2} d_{g_3} \nu^3 \langle \Delta_R \rangle_0$. For nucleon decay, considering the available phase space, the relevant generational indices for $u_g$ is $g=1$. Similarly for $d_g$ we can have $g=1,2$, thus allowing a strange (anti)quark in the final state, which can then form a kaon. Evidently the operator can mediate three-body nucleon decays such as $p \to \ell^+ {\bar \nu}^\prime {\bar \nu}^{\prime \prime}$ and $n \to {\bar \nu} {\bar \nu^\prime} {\bar \nu^{\prime \prime}}$. Taking into account spectator quarks in the nucleon, this same six-fermion BNV operator will mediate decays involving three (anti)leptons and a meson, although these may be phase-space suppressed. These include decays such as $p \to \ell^+ {\bar \nu^\prime} {\bar \nu^{\prime \prime} \pi^0}$, $p \to \ell^+ {\bar \nu^\prime} {\bar \nu^{\prime \prime} K^0}$, $n \to \ell^+ {\bar \nu^\prime} {\bar \nu^{\prime \prime} \pi^-}$, $p \to {\bar \nu} {\bar \nu^\prime} {\bar \nu^{\prime \prime} \pi^+}$, $p \to {\bar \nu} {\bar \nu^\prime} {\bar \nu^{\prime \prime} K^+}$, $n \to {\bar \nu} {\bar \nu^\prime} {\bar \nu^{\prime \prime} \pi^0}$, $n \to {\bar \nu} {\bar \nu^\prime} {\bar \nu^{\prime \prime} K^0}$, etc. (see Fig. \ref{Nm3_figure}).

As this demonstrates, each operator can mediate several different decays, and each class contains several operators. The theoretical framework of the LRS model in extra dimensions enables us to analyze them in the same way. The LRS model specifies the mass scale responsible for these $|\Delta (B-L)| = 2$ processes, namely $v_R$, and each class determines the exponential suppression factor coming from the integration of fermion wave functions over the extra dimensions. Using eq. \eqref{intform}, we write down the suppression factors, ${\cal  I}_{C_r^{(Nm3)}} \equiv e^{-S_r^{(Nm3)}}$ for each class (see Table \ref{Nm3_class_table}):
%
\begin{align}
        {\cal  I}_{C_1^{(Nm3)}}  ={}& \exp \Big[-\frac{3}{2} \|\eta_{Q_R}-\eta_{L_R}\|^2\Big]  \ , \label{I1_Nm3_eqn} \\[7pt] 
        {\cal  I}_{C_2^{(Nm3)}} ={}& \exp \Big[-\frac{1}{6}\Big\{ 6 \|\eta_{Q_R}-\eta_{L_L}\|^2 + 3 \|\eta_{Q_R}-\eta_{L_R}\|^2 + 2 \|\eta_{L_L}-\eta_{L_R}\|^2 \Big\}\Big] \ , \label{I2_Nm3_eqn}  \\[7pt]
        \begin{split}
                   {\cal  I}_{C_3^{(Nm3)}}  ={}&  \exp \Big[-\frac{1}{6}\Big\{2 \|\eta_{Q_L}-\eta_{Q_R}\|^2 + \|\eta_{Q_L}-\eta_{L_L}\|^2 + 2 \|\eta_{Q_L}-\eta_{L_R}\|^2 + \\ 
                   & + 2 \|\eta_{Q_R}-\eta_{L_L}\|^2 + 4 \|\eta_{Q_R}-\eta_{L_R}\|^2 + 2 \|\eta_{L_L}-\eta_{L_R}\|^2 \Big\}\Big]
                     \ ,  \label{I3_Nm3_eqn}
        \end{split} \\[7pt] 
        {\cal  I}_{C_4^{(Nm3)}} ={}& \exp \Big[-\frac{1}{6}\Big\{ 2 \|\eta_{Q_L}-\eta_{Q_R}\|^2 + 6 \|\eta_{Q_L}-\eta_{L_R}\|^2 + 3 \|\eta_{Q_R}-\eta_{L_R}\|^2 \Big\}\Big] \ , \label{I4_Nm3_eqn}  \\[7pt]
       \begin{split}
                    {\cal  I}_{C_5^{(Nm3)}}  ={}&  \exp \Big[-\frac{1}{6}\Big\{2 \|\eta_{Q_L}-\eta_{Q_R}\|^2 + 4 \|\eta_{Q_L}-\eta_{L_L}\|^2 + 2 \|\eta_{Q_L}-\eta_{L_R}\|^2 + \\ 
                    & + 2 \|\eta_{Q_R}-\eta_{L_L}\|^2 + \|\eta_{Q_R}-\eta_{L_R}\|^2 + 2 \|\eta_{L_L}-\eta_{L_R}\|^2 \Big\}\Big] \ ,
                    \label{I5_Nm3_eqn}   
       \end{split} \\[7pt]
        {\cal  I}_{C_6^{(Nm3)}} ={}& \exp \Big[-\frac{1}{6}\Big\{ 2 \|\eta_{Q_L}-\eta_{Q_R}\|^2 + 3 \|\eta_{Q_L}-\eta_{L_L}\|^2 + 6 \|\eta_{Q_R}-\eta_{L_L}\|^2 \Big\}\Big] \ , \label{I6_Nm3_eqn}  \\[7pt]
        {\cal  I}_{C_7^{(Nm3)}} ={}& \exp \Big[-\frac{1}{6}\Big\{ 6 \|\eta_{Q_L}-\eta_{L_R}\|^2 + 3 \|\eta_{Q_L}-\eta_{L_L}\|^2 + 2 \|\eta_{L_L}-\eta_{L_R}\|^2 \Big\}\Big]  \ . \label{I7_Nm3_eqn}
    \end{align}

With the knowledge of the exponential suppression factors arising from the integration of fermion wave functions over the extra dimensions, let us proceed to estimate the decay rates of $Nm3$ processes. As the four-body decay modes are suppressed by phase-space, we focus on the three-body decays. As the daughter particles are very light compared with $m_N$, using eq. \eqref{Gamma_special_eqn} with $n_f = 6$, $M_{{\cal X}} = M_{Nm3} = v_R$, $m_A = m_N$ and $P=3$ we get
\beq
      \Gamma_{Nm3} \sim{} m_N \,  \frac{S}{2^9 \pi^3} \, \bigg(\frac{\Lambda_{QCD}}{v_R}\bigg)^{10} \, \bigg(\frac{2^{1/2} \, \mu}{3^{1/4} \, \pi^{1/2} \, v_R}\bigg)^{4n}
       \, \bigg|\sum_r {\bar\kappa}_r^{(Nm3)} \, e^{-S_r^{(Nm3)}}\bigg|^2 \ ,
        \label{Gamma_Nm3_eqn} 
\eeq
    where $S$ is the symmetry factor if there are identical final-state particles, and $\bar \kappa_r^{(Nm3)}$ are ${\cal O}(1)$ dimensionless coefficients. 
    
Four-fermion operators mediating nucleon decay constrain the separation of quark and lepton wave functions in the extra dimensions. Furthermore, the experimental bound on $n$-$\bar n$ oscillations implies a lower bound on the $B-L$ breaking scale, $v_R$. We obtained those bounds in ref. \cite{nnblrs} (see eqs. (4.17), (5.19) in \cite{nnblrs}). Using these above-mentioned bounds in eq. \eqref{Gamma_Nm3_eqn}, we see that the decay rates for these $Nm3$ decays are suppressed well below current experimental bounds. These experimental limits on the partial lifetimes for $Nm3$ decay modes include $(\tau/B)_{p \to e^+xx} > 1.7 \times 10^{32}$ yr.\ \cite{takhistov_ptolxx}, $(\tau/B)_{p \to \mu^+xx} > 2.2 \times 10^{32}$ yr.\ \cite{takhistov_ptolxx}, and $(\tau/B)_{n \to xxx} > 5.8 \times 10^{29}$ yr.\ \cite{Akamland_nn_to_nunubar}, where $x$ denotes a weakly interacting fermion having negligible mass which does not decay in the detector.
\section{Dinucleon decays to dileptons with \boldmath $\Delta B=-2$ and \boldmath $\Delta L=-2$}
\label{dinucleon_section}

In this section we consider, in the framework of the LRS extra-dimensional model, dinucleon decays to dileptons with $\Delta B=-2$ and $\Delta L=-2$, such as $pp \to \ell^+ \ell^{\prime +}$, $np \to \ell^+ {\bar\nu}$ and $nn \to {\bar\nu} {\bar\nu^\prime}$. These decays thus respect the $B-L$ symmetry. The same operators mediating $n$-$\bar n$ oscillations, coupled with standard model $W$, $Z$ vector bosons, can mediate processes like $np \to \ell^+ \nu$, $nn \to \bar \nu \nu$ and $nn \to \ell^+ \ell^-$, where $\ell$ is $e$, $\mu$. These processes have $\Delta B = -2$ and $\Delta L = 0$, thereby violating $B-L$ by two units. The operators mediating these dinucleon-to-dilepton decays are comprised of eight fermion fields.  Ref. \cite{dnd} derived estimates for upper bounds on rates for these decay modes in a general phenomenological framework (not connected with the present extra-dimensional model).  Here, in contrast, we will focus on the $(B-L)$-conserving decay modes.  We remark that the decay $pp \to e^+ e^+$ is related to hydrogen-antihydrogen oscillations \cite{HHb78,HHb_mohapatra} by charge conjugation symmetry. We denote these decays by the notation $(N N^\prime)$.

Proceeding with the analysis, we will first derive the different classes mediating $(N N^\prime)$ decays. As these decays conserve $B-L$, the leading operators that mediate them do not involve (VEVs of) $\Delta_R$ fields. The lowest-dimensional operator classes contributing to these decays thus have the schematic form
\beq \label{NN_op_eqn}
{C}_r^{(N N^\prime)} \sim{} Q_L^{n_{Q_L}} \, Q_R^{n_{Q_R}} \, L_L^{n_{L_L}} \, L_R^{n_{L_R}} \ .
\eeq
The exponent of a certain field indicates the number of that field present in that class. As we are considering dinucleon initial states and dilepton final state, we require
\begin{align}
    n_{Q_L} + n_{Q_R} ={} & 6 \label{nq_eqn_NN} \\
    n_{L_L} + n_{L_R} ={}& 2 \label{nL_eqn_NN} \ .
\end{align}
The condition of $U(1)_{B-L}$ invariance is
\beq \label{B-L_eqn_NN}
    \frac{1}{3} (n_{Q_L}+n_{Q_R}) - (n_{L_L}+n_{L_R}) = 0 \ .
\eeq
The combination of eq. \eqref{nq_eqn_NN} and eq. \eqref{nL_eqn_NN} automatically satisfies eq. \eqref{B-L_eqn_NN}. The triality condition for $SU(3)_c$ invariance is automatically satisfied by eq. \eqref{nq_eqn_NN} as $\frac{1}{3} (n_{Q_L}+ n_{Q_R}) = 2 \in {\mathbb N}$. The respective requirements that the operators must be invariant under ${\rm SU}(2)_L$ and ${\rm SU}(2)_R$ imply that 
\begin{align}
    {\widetilde{N}}_L \equiv n_{Q_L} + n_{L_L} ={} & {\mathrm{even}} \label{SU2L_eqn_NN} \\
    {\widetilde{N}}_R \equiv  n_{Q_R} + n_{L_R} ={}& {\mathrm{even}} \label{SU2R_eqn_NN} \ .
\end{align}
If there is an even number of derivatives present in an operator belonging to a certain class, then the condition of Lorentz invariance is identical with eqs. \eqref{SU2L_eqn_NN}, \eqref{SU2R_eqn_NN}. On the other hand, if there were an odd number of derivatives present, then Lorentz invariance would be violated by
eqs. \eqref{SU2L_eqn_NN}, \eqref{SU2R_eqn_NN}.  Therefore, only an even number of derivatives is possible in this case. For lowest-dimensional operators, we choose the number of derivatives to be zero. Clearly eqs. \eqref{nq_eqn_NN}, \eqref{nL_eqn_NN}, \eqref{SU2L_eqn_NN} contain eq. \eqref{SU2R_eqn_NN}. Looking for non-negative integer solutions for the linearly independent equations, namely eqs. \eqref{nq_eqn_NN}, \eqref{nL_eqn_NN}, and \eqref{SU2L_eqn_NN}, we obtain the classes of operators. These classes are listed on Table \ref{NN_class_table}.
\begingroup
\setlength{\tabcolsep}{10pt} 
\renewcommand{\arraystretch}{1.5} 
    \begin{table}[ht]
        \begin{center}
            \begin{tabular}{|c|c|c|c|} \hline\hline
Class $C_{r}^{(N N^\prime)}$   & $\widetilde{N}_L$  & $\widetilde{N}_R$  & Structure   \\ \hline
                    $C_1^{(N N^\prime)}$ & $0$ & $8$ & $Q_R^6 \, L_R^2$ \\[6pt]
                    $C_2^{(N N^\prime)}$ & $2$ & $6$ & $Q_R^6 \, L_L^2$ \\[6pt]
                    $C_3^{(N N^\prime)}$ & $2$ & $6$ & $Q_L \, Q_R^5 \, L_L \, L_R$ \\[6pt]
                    $C_4^{(N N^\prime)}$ & $2$ & $6$ & $Q_L^2 \, Q_R^4 \, L_R^2$ \\[6pt]
                    $C_5^{(N N^\prime)}$ & $4$ & $4$ & $Q_L^2 \, Q_R^4 \, L_L^2$ \\[6pt]
                    $C_6^{(N N^\prime)}$ & $4$ & $4$ & $Q_L^4 \, Q_R^2  \, L_R^2$ \\[6pt]
                    $C_7^{(N N^\prime)}$ & $4$ & $4$ & $Q_L^3 \, Q_R^3 \, L_L \, L_R$ \\[6pt]
                    $C_8^{(N N^\prime)}$ & $6$ & $2$ & $Q_L^6 \, L_R^2$ \\[6pt]
                    $C_9^{(N N^\prime)}$ & $6$ & $2$ & $Q_L^5 \, Q_R \, L_L \, L_R$ \\[6pt]
                    $C_{10}^{(N N^\prime)}$ & $6$ & $2$ & $Q_L^4 \, Q_R^2 \, L_L^2$ \\[6pt]
                     $C_{11}^{(N N^\prime)}$ & $8$ & $0$ & $Q_L^6 \, L_L^2$ \\[6pt]
            \hline\hline
            \end{tabular}
        \end{center}
            \caption{List of classes of operators mediating $\Delta B = -2$, $\Delta L = -2$ $N N^\prime$ processes such as $pp \to \ell^+ \ell^{\prime +}$, $np \to \ell^+ \bar\nu$, $nn \to \bar\nu \bar\nu$. The numbers $\widetilde{N}_L$ and $\widetilde{N}_R$ denote the total numbers of left-handed 
            and right-handed fermion doublets, respectively, that are present in a class.}
    \label{NN_class_table}
\end{table}
\endgroup

Clearly, the classes come in left-right-symmetric pair, reflecting the $(B-L)$-conserving nature of the $N N^\prime$ processes, in contrast with Table \ref{Nm3_class_table}. Taking into account spectator quarks in the nucleons, the $(N N^\prime)$ operators can also give rise to dinucleon decays to three-body final states with $\Delta B=-2$ and $\Delta L =-2$. For example, these include $N p \to \ell^+ \ell^{\prime +} M$, $N p \to \ell^+ \bar \nu M$, $nn \to \ell^+ \bar \nu M^-$, and $N n \to \bar\nu \bar\nu M$, where the notation subsumes multiple processes by denoting $n,p$ with $N$ and a light meson allowed by phase space by $M$, with the appropriate electric charge required for charge conservation in the process. To illustrate, the notation $N p \to \ell^+ \bar \nu M$ represents $pp \to \ell^+ \bar \nu \pi^+$, $pp \to \ell^+ \bar \nu K^+$, $np \to \ell^+ \bar \nu \pi^0$, $np \to \ell^+ \bar\nu K^0$ and so forth. As these operators are  related to the same operators that mediate the two-body decays having the same selection rule, and as these three-body modes are also phase-space suppressed, estimating the rates for the two-body $N N^\prime$ decays is sufficient to check if all these processes are in accord with experimental limits. To this end, using eq. \eqref{intform}, we write down the suppression factor originating from the extra-dimensional integration ${\cal I}_{r}^{(N N^\prime)} \equiv e^{-S_r^{(N N^\prime)}}$ for each class:
        \begin{align}
        {\cal  I}_{C_1^{(N N^\prime)}}  ={}& \exp \Big[-\frac{3}{2} \|\eta_{Q_R}-\eta_{L_R}\|^2\Big]  \ , \label{I1_NN_eqn} \\[7pt]
        {\cal  I}_{C_2^{(N N^\prime)}}  ={}& \exp \Big[-\frac{3}{2} \|\eta_{Q_R}-\eta_{L_L}\|^2\Big]  \ , \label{I2_NN_eqn} \\[7pt]
    \begin{split}
         {\cal  I}_{C_3^{(N N^\prime)}} ={}& \exp \Big[-\frac{1}{8}\Big\{ 5 \|\eta_{Q_L}-\eta_{Q_R}\|^2 +  \|\eta_{Q_L}-\eta_{L_L}\|^2 +  \|\eta_{Q_L}-\eta_{L_R}\|^2 + \\ & + 5 \|\eta_{Q_R}-\eta_{L_L}\|^2 + 5 \|\eta_{Q_R}-\eta_{L_R}\|^2 + \|\eta_{L_L}-\eta_{L_R}\|^2\Big\}\Big] \ , \label{I3_NN_eqn} 
         \end{split} \\[7pt]
           {\cal  I}_{C_4^{(N N^\prime)}} ={}& \exp \Big[-\frac{1}{2}\Big\{ 2 \|\eta_{Q_L}-\eta_{Q_R}\|^2 +  \|\eta_{Q_L}-\eta_{L_R}\|^2 + 2 \|\eta_{Q_R}-\eta_{L_R}\|^2 \Big\}\Big] \ , \label{I4_NN_eqn} \\[7pt]
         {\cal{I}}_{C_5^{(N N^\prime)}} ={}& \exp \Big[-\frac{1}{2}\Big\{ 2 \|\eta_{Q_L}-\eta_{Q_R}\|^2 +  \|\eta_{Q_L}-\eta_{L_L}\|^2 + 2 \|\eta_{Q_R}-\eta_{L_L}\|^2 \Big\}\Big] \ , \label{I5_NN_eqn} 
         \end{align}
         \begin{align}
        {\cal{I}}_{C_6^{(N N^\prime)}} ={}& \exp \Big[-\frac{1}{2}\Big\{ 2 \|\eta_{Q_L}-\eta_{Q_R}\|^2 + 2 \|\eta_{Q_L}-\eta_{L_R}\|^2 +  \|\eta_{Q_R}-\eta_{L_R}\|^2 \Big\}\Big]  \ , \label{I6_NN_eqn} \\[7pt]
        \begin{split}
                {\cal  I}_{C_7^{(N N^\prime)}} ={}& \exp \Big[-\frac{1}{8}\Big\{ 6 \|\eta_{Q_L}-\eta_{Q_R}\|^2 + 3 \|\eta_{Q_L}-\eta_{L_L}\|^2 + 3 \|\eta_{Q_L}-\eta_{L_R}\|^2 + \\ & + 3 \|\eta_{Q_R}-\eta_{L_L}\|^2 + 3 \|\eta_{Q_R}-\eta_{L_R}\|^2 + \|\eta_{L_L}-\eta_{L_R}\|^2\Big\}\Big]  \ , \label{I7_NN_eqn}
                \end{split} \\[7pt]
        {\cal  I}_{C_8^{(N N^\prime)}}  ={}& \exp \Big[-\frac{3}{2} \|\eta_{Q_L}-\eta_{L_R}\|^2\Big]  \ , \label{I8_NN_eqn} \\[7pt]
        \begin{split}
                    {\cal  I}_{C_9^{(N N^\prime)}} ={}& \exp \Big[-\frac{1}{8}\Big\{ 5 \|\eta_{Q_L}-\eta_{Q_R}\|^2 + 5 \|\eta_{Q_L}-\eta_{L_L}\|^2 + 5 \|\eta_{Q_L}-\eta_{L_R}\|^2 + \\ &+  \|\eta_{Q_R}-\eta_{L_L}\|^2 +  \|\eta_{Q_R}-\eta_{L_R}\|^2 + \|\eta_{L_L}-\eta_{L_R}\|^2\Big\}\Big] \ , \label{I9_NN_eqn}
                    \end{split} \\[7pt] 
        {\cal{I}}_{C_{10}^{(N N^\prime)}} ={}& \exp \Big[-\frac{1}{2}\Big\{ 2 \|\eta_{Q_L}-\eta_{Q_R}\|^2 + 2 \|\eta_{Q_L}-\eta_{L_L}\|^2 +  \|\eta_{Q_R}-\eta_{L_L}\|^2 \Big\}\Big] \ , \label{I10_NN_eqn} \\[7pt]
        {\cal  I}_{C_{11}^{(N N^\prime)}}  ={}& \exp \Big[-\frac{3}{2} \|\eta_{Q_L}-\eta_{L_L}\|^2\Big]  \ . \label{I11_NN_eqn}
        \end{align}

Considering the two body decays, where the outgoing particles are very light compared with $2 m_N$, we can use eq. \eqref{Gamma_special_eqn} to estimate the decay rate. Using $n_f = 8$, $n_{\Delta_R} = 0$, $P=2$, $m_A = 2 m_N$, and $M_{\cal X} = M_{N N^\prime}$, we estimate:
\beq
      \Gamma_{N N^\prime} \sim{}  \frac{S}{2^5 \pi \, m_N} \, \frac{(\Lambda_{QCD})^{18}}{M_{N N^\prime}^{16}} \, \bigg(\frac{2^{1/6} \, \mu}{\pi^{1/2} \, M_{N N^\prime}}\bigg)^{6n}
       \, \bigg|\sum_r {\bar\kappa}_r^{(N N^\prime)} \, e^{-S_r^{(N N^\prime)}}\bigg|^2 \ .
        \label{Gamma_NN_eqn} 
\eeq
As before, $S$ is a symmetry factor, and $M_{N N^\prime}$ is the mass scale characterizing the $N N^\prime$ processes. All the fermion wave function separation distances appearing in the exponential suppression factors have lower bound constraints from the nucleon decay lifetime limits (see eq. (4.17) in ref. \cite{nnblrs}). Using this constraint together with the higher power of $1/M_{N N^\prime}$ and the exponential suppression factors, we find that the $N N^\prime$ decay rates are well below current experimental limits. These experimental bounds on the partial lifetimes for $NN^\prime$ decay modes include $(\tau/B)_{pp \to e^+ e^+} > 5.8 \times 10^{30}$ yr.\ \cite{frejus}, $(\tau/B)_{pp \to e^+ \mu^+} > 3.6 \times 10^{30}$ yr.\ \cite{frejus}, $(\tau/B)_{pp \to \mu^+ \mu^+} > 1.7 \times 10^{30}$ yr.\ \cite{frejus}; $(\tau/B)_{np \to e^+ x} > 2.6 \times 10^{32}$ yr.\ \cite{takhistov15}, $(\tau/B)_{np \to \mu^+ x} > 2.0 \times 10^{32}$ yr.\ \cite{takhistov15}, $(\tau/B)_{np \to \tau^+ x} > 2.9 \times 10^{31}$ yr.\ \cite{takhistov15}, and $(\tau/B)_{nn \to x x} > 1.4 \times 10^{30}$ yr.\ \cite{Akamland_nn_to_nunubar}. $x$ stands for a weakly interacting fermion having negligible mass which does not decay in the detector.

One can consider other BNV decay modes and perform a similar analysis. In the LRS model, all of the four possible quark-lepton wave function separation distances $\|\eta_{Q_L}-\eta_{L_L}\|$,
$\|\eta_{Q_L}-\eta_{L_R}\|$,
$\|\eta_{Q_R}-\eta_{L_L}\|$, and
$\|\eta_{Q_R}-\eta_{L_R}\|$ are constrained from limits on nucleon decay modes mediated by four-fermion operators (see eq. (4.17) in ref. \cite{nnblrs}). It is clear from eq. \eqref{intform} that all pairs of fermion wave function separation distances enter in the exponential suppression factors $e^{-S_r}$. Consequently, in general, any class of operators which is comprised of both quarks and leptons will be suppressed exponentially. In contrast, $n$-$\bar n$ oscillations contain only quark wave functions, and ref. \cite{nnblrs} found that some of the classes of six-quark operators in the LRS model have no suppression from the integration over extra dimensions.  
\section{Conclusions}
\label{conclusions_section}

There is strong motivation to investigate baryon-number-violating processes. 
Ref. \cite{nnb02} found the interesting result, using a Standard Model effective field theory framework in a model with large extra dimensions, that although one can suppress nucleon decay well below experimental limits by separating quark and lepton wave functions in the extra dimensions, this does not suppress $n-\bar n$ oscillations, which can occur at levels comparable to experimental bounds. Recently, in ref. \cite{nnblrs} we found, in the context of a left-right-symmetric model 
with large extra dimensions, that this effect is even more pronounced, because some classes of six-quark operators mediating $n-\bar n$ transitions do not incur any exponential suppression factors from the integration of the wave functions over the extra dimensions. Ref. \cite{nnblrs} used this to derive a lower bound on the left-right symmetry breaking scale, $v_R$, in this extra-dimensional model. Motivated by these findings, in this paper we analyzed (a) nucleon decays to trileptons with $\Delta B=-1$ and $\Delta L=-3$ and (b) dinucleon decays to dileptons with $\Delta B=-2$ and $\Delta L=-2$ in the extra-dimensional left-right-symmetric model. Assuming the minimal field content in table \ref{LRS_field_content_table}, we classified the operators into different classes having different suppression factors resulting from the integrations over the extra dimensions to derive the low-energy effective Lagrangians. Using a general approach as in refs. \cite{bvd,nnblrs}, we searched for all of the classes of operators for these decays and obtained the different exponential suppression factors for each class. Estimation of the decay rate for these processes has been given. Using the bounds on $v_R$ and on quark-lepton wave function separations in the higher dimensions \cite{nnblrs}, we have demonstrated that these modes are suppressed sufficiently to be in accord with experimental limits. 

\acknowledgments
I would like to thank Prof.\ R.\ Shrock for helpful discussions. This research was partly supported by NSF Grant NSF-PHY-1915093. 

\appendix
\section{An integration formula} \label{intform_appendix}
The following integration formula from ref. \cite{bvd} is restated here for the convenience of the reader. This determines the exponential suppression factor once a class is specified. 
\beq
\int_{-\infty}^\infty d^n \eta \, \exp
\Big [-\sum_{i=1}^m a_i\|\eta-\eta_{f_i}\|^2 \Big ]
= \bigg [ \frac{\pi}{\sum_{i=1}^m a_i} \bigg ]^{n/2} \,
\exp\Bigg [ \frac{-\sum_{j,k=1; \ j < k}^m \, a_j a_k
\|\eta_{f_j}-\eta_{f_k}\|^2}{\sum_{s=1}^m a_s} \Bigg ] \ .
\label{intform}
\eeq
\section{Decay width} \label{decay_width_appendix}
Consider a BNV decay process $A \to \mathrm{f.s.}$, where an initial state $A$ (a nucleon or dinucleon in our applications) with a mass $m_A$ decays to a $P$-body final state $\mathrm{f.s.}$. Let us denote this process by ${\cal X}$ and corresponding effective mass scale characterizing the physics responsible for this decay as $M_{\cal X}$. Eq. \eqref{Leff_4dim_eqn} represents the Lagrangian responsible for this decay, in which the operators mediating the decay consist of $n_f$ fermion fields and $n_{\Delta_R}$ triplet Higgs fields. The decay rate is 
\beq \label{Decay_width_eqn}
\Gamma_{\cal X} ={} \frac{S}{2 m_A} \int dR_P \, \Big|\sum_r c_r^{({\cal X})} \langle \mathrel{f.s.}|{\cal O}_r^{({\cal X})}|A \rangle \Big|^2 \ .
\eeq
The average over spin and polarization is implicit. $S$ is the symmetry factor to take into account possible identical particles in $\mathrel{f.s.}$; $dR_P$ represents the $P$-body phase space integration and is defined as follows:
\beq \label{R_P_definition_eqn}
    \int dR_P \equiv \frac{1}{(2\pi)^{3P-4}} \,
\int \Big [ \prod_{i=1}^P 
\frac{d^3 k_i}{2E_i} \Big ] \, \delta^4 \Big ( k-(\sum_{i=1}^P k_i) \Big ) \ ,
\eeq
where $k_i$ denotes momentum of the $i^{\mathrm{th}}$ outgoing particle. One can also consider the phase space by itself: $R_P = \int dR_P$. For the case where the outgoing particles have negligible mass in comparison with the mass of the parent particle ($m_A$), it is useful to define a dimensionless phase space factor
\beq \label{dimensionless_R_P_eqn}
    {\bar R}_{P , \, 0} \equiv{} (m_A)^{-(2P-4)} R_P = \frac{1}{2^{4P-5} \pi^{2P-3} \Gamma(P) \Gamma(P-1)} \quad {\rm for} \ P \ge 2  \ .
\eeq
The $P$-body phase space integration for $P \geq 3$ is non-trivial, and requires full knowledge of the $c_r^{(\cal X)}$ coefficients. But the values of these coefficients are not specified in the framework of the low-energy effective field theory, since they depend on the ultraviolet completion. Hence, we only give an estimate for the decay rate by taking the phase space into account by the factor $R_P$. As $\langle \mathrel{f.s.}|{\cal O}_r^{({\cal X})}|A \rangle$ is determined from QCD, there is only one relevant scale for this matrix element, namely $\Lambda_{QCD}$ ($\sim 0.25$ GeV). So this matrix element can be approximated by $(\Lambda_{QCD})^m$, where from dimensional analysis $m = 3 n_f/2 - (1+P)$. Hence from eqs. \eqref{c_kappa_relation_eqn}, \eqref{Decay_width_eqn}
\begin{align}
    \begin{split}
    \Gamma_{{\cal X}} \sim {}& \frac{S}{2 m_A} \, R_P \, \frac{1}{M_{{\cal X}}^{(3 n_f - 8)}} \, \bigg(\frac{v_R}{M_{{\cal X}}}\bigg)^{2 n_{\Delta_R}} \, \bigg(\frac{\mu}{M_{{\cal X}}}\bigg)^{(n_f - 2)n} \, \bigg(\frac{2^{n_f/2}}{\pi^{(n_f-2)/2} \, n_f}\bigg)^n \times \\
    & \times \bigg|\sum_r {\bar \kappa}_r^{({\cal X})} \, e^{-S_r^{({\cal X})}}\bigg|^2 \, \Lambda_{QCD}^{(3 n_f - 2 (1+P))} \ .
   \label{Gamma_general_eqn} 
   \end{split}
\end{align}
Specializing for the case when the masses of the daughter particles are negligible in comparison with $m_N$, using eq. \eqref{dimensionless_R_P_eqn}, we get
\begin{equation}
    \begin{split}
        \Gamma_{{\cal X}} \sim {} & \frac{S}{2^{4(P-1)}\pi^{2P-3} \Gamma(P) \Gamma(P-1)} \, \frac{m_A^{(2P-5)}}{M_{{\cal X}}^{(3 n_f - 8)}} \, \bigg(\frac{v_R}{M_{{\cal X}}}\bigg)^{2 n_{\Delta_R}} \, \bigg(\frac{\mu}{M_{{\cal X}}}\bigg)^{(n_f - 2)n} \times \\
        & \times \bigg(\frac{2^{n_f/2}}{\pi^{(n_f-2)/2} \, n_f}\bigg)^n \,  \bigg|\sum_r {\bar \kappa}_r^{({\cal X})} \, e^{-S_r^{({\cal X})}}\bigg|^2 \, \Lambda_{QCD}^{(3 n_f - 2 (1+P))} \ .
        \label{Gamma_special_eqn}
    \end{split}
\end{equation}
\section{Explicit operators in each class} \label{operators_Nm3_appendix}
Although the analysis in the text does not require explicit operators, it is useful to construct some explicit operators belonging to each class, as different operators contribute to different processes. 

\subsection{Operators for nucleon to trilepton decays}
\label{Nm3_appendix_subsection}

We list here examples of operators belonging to each class mediating nucleon decays to trileptons (see Table \ref{Nm3_class_table}). The superscript on an operator denotes the three-body decays that it mediates. For example, the notation ${\cal O}_2^{(nm3,pm3)}$ indicates that this operator mediates both $p \to \ell^+ \bar\nu \bar\nu^\prime$ and $n \to \bar\nu \bar\nu^\prime \bar\nu^{\prime\prime}$ processes. As the classes are manifestly independent of each other, it is not required to analyze the linear independence of the explicit operators. The generational indices on fermion fields are suppressed in the notation.
    \begin{multline}
       {\mathcal{O}}_2^{(nm3,pm3)} = {} \big[Q_{R}^{i^\prime \, \alpha  T} C L_{R}^{j^\prime}\big] \, \big[Q_{R}^{k^\prime \, \beta \, T} C L_{R}^{m^\prime}\big] \, \big[Q_{R}^{p^\prime \, \gamma  T} C L_{R}^{q^\prime}\big] \, (\Delta_R)^{r^\prime s^\prime} \times \\  \times \epsilon_{\alpha \beta \gamma} \, \epsilon_{i^\prime j^\prime} \, \epsilon_{k^\prime m^\prime} \, (\epsilon_{p^\prime r^\prime} \, \epsilon_{q^\prime s^\prime}+\epsilon_{p^\prime s^\prime} \, \epsilon_{q^\prime r^\prime}) \in C_1^{(Nm3)} \ .
       \label{O2_Nm3_eqn}
    \end{multline}
    \begin{multline}
       {\mathcal{O}}_3^{(pm3)} = {} \big[Q_{R}^{i^\prime \, \alpha  T} C Q_{R}^{j^\prime \beta}  \big] \, \big[Q_{R}^{k^\prime \, \gamma \, T} C L_{R}^{m^\prime}\big] \, \big[L_{L}^{p \, T} C L_{L}^{q}\big] \, (\Delta_R)^{r^\prime s^\prime} \times \\  \times \epsilon_{\alpha \beta \gamma} \, \epsilon_{i^\prime j^\prime} \, \epsilon_{p q} \, (\epsilon_{k^\prime r^\prime} \, \epsilon_{m^\prime s^\prime}+\epsilon_{k^\prime s^\prime} \, \epsilon_{m^\prime r^\prime}) \in C_2^{(Nm3)} \ .
       \label{O3_Nm3_eqn}
    \end{multline}
      \begin{multline}
       {\mathcal{O}}_4^{(pm3,nm3)} = {} \big[Q_{R}^{i^\prime \, \alpha  T} C Q_{R}^{j^\prime \beta}  \big] \, \big[Q_{L}^{k \, \gamma \, T} C L_{L}^{m}\big] \, \big[L_{R}^{p^\prime \, T} C L_{R}^{q^\prime}\big] \, (\Delta_R)^{r^\prime s^\prime} \times \\  \times \epsilon_{\alpha \beta \gamma} \, \epsilon_{i^\prime j^\prime} \, \epsilon_{k m} \, (\epsilon_{p^\prime r^\prime} \, \epsilon_{q^\prime s^\prime}+\epsilon_{p^\prime s^\prime} \, \epsilon_{q^\prime r^\prime}) \in C_3^{(Nm3)} \ .
       \label{O4_Nm3_eqn}
    \end{multline}
      \begin{multline}
       {\mathcal{O}}_5^{(pm3,nm3)} = {} \big[Q_{L}^{i \, \alpha  T} C Q_{L}^{j \beta}  \big] \, \big[Q_{R}^{k^\prime \, \gamma \, T} C L_{R}^{m^\prime}\big] \, \big[L_{R}^{p^\prime \, T} C L_{R}^{q^\prime}\big] \, (\Delta_R)^{r^\prime s^\prime} \times \\  \times \epsilon_{\alpha \beta \gamma} \, \epsilon_{i j} \, \epsilon_{k^\prime m^\prime} \, (\epsilon_{p^\prime r^\prime} \, \epsilon_{q^\prime s^\prime}+\epsilon_{p^\prime s^\prime} \, \epsilon_{q^\prime r^\prime}) \in C_4^{(Nm3)} \ .
       \label{O5_Nm3_eqn}
    \end{multline}
          \begin{multline}
       {\mathcal{O}}_6^{(pm3)} = {} \big[Q_{L}^{i \, \alpha  T} C Q_{L}^{j \beta}  \big] \, \big[Q_{R}^{k^\prime \, \gamma \, T} C L_{R}^{m^\prime}\big] \, \big[L_{L}^{p \, T} C L_{L}^{q}\big] \, (\Delta_R)^{r^\prime s^\prime} \times \\  \times \epsilon_{\alpha \beta \gamma} \, \epsilon_{i j} \, \epsilon_{p q} \, (\epsilon_{k^\prime r^\prime} \, \epsilon_{m^\prime s^\prime}+\epsilon_{k^\prime s^\prime} \, \epsilon_{m^\prime r^\prime}) \in C_5^{(Nm3)} \ .
       \label{O6_Nm3_eqn}
    \end{multline}
          \begin{multline}
       {\mathcal{O}}_7^{(pm3)} = {} \big[Q_{R}^{i^\prime \, \alpha  T} C Q_{R}^{j^\prime \beta}  \big] \, \big[Q_{L}^{k \, \gamma \, T} C L_{L}^{m}\big] \, \big[L_{L}^{p \, T} C L_{L}^{q}\big] \, (\Delta_R)^{r^\prime s^\prime} \times \\  \times \epsilon_{\alpha \beta \gamma} \, \epsilon_{k m} \, \epsilon_{p q} \, (\epsilon_{i^\prime r^\prime} \, \epsilon_{j^\prime s^\prime}+\epsilon_{i^\prime s^\prime} \, \epsilon_{j^\prime r^\prime}) \in C_6^{(Nm3)} \ .
       \label{O7_Nm3_eqn}
    \end{multline}
        \begin{multline}
       {\mathcal{O}}_8^{(pm3,nm3)} = {} \big[Q_{L}^{i \, \alpha  T} C Q_{L}^{j \beta}  \big] \, \big[Q_{L}^{k \, \gamma \, T} C L_{L}^{m}\big] \, \big[L_{R}^{p^\prime \, T} C L_{R}^{q^\prime}\big] \, (\Delta_R)^{r^\prime s^\prime} \times \\  \times \epsilon_{\alpha \beta \gamma} \, \epsilon_{i j} \, \epsilon_{k m} \, (\epsilon_{p^\prime r^\prime} \, \epsilon_{q^\prime s^\prime}+\epsilon_{p^\prime s^\prime} \, \epsilon_{q^\prime r^\prime}) \in C_7^{(Nm3)} \ .
       \label{O8_Nm3_eqn}
    \end{multline}
    
We note that operator ${\cal O}_{7}^{(pm3)}$ in eq. \eqref{O7_Nm3_eqn} vanishes, due to the fact that $Q_R$ fields are contracted anti-symmetrically in the color index, but symmetrically in the weak $SU(2)_R$ index. If the $Q_R$ fields are of the same generation, then Fermi statistics renders this operator zero. Explicitly, as only the neutral component of $\Delta_R$ picks up a VEV, writing the generational indices explicitly, it is easy to see that the $(u_{R,{g_1}}^{\alpha T} C u_{R,g_2}^\beta)$ combination is selected from $(Q_{R,g_1}^{i^\prime \alpha T} \, C \, Q_{R,g_2}^{j^\prime \beta})$ part, due to charge conservation. However, all the $u$ fields have to be from the first generation, because single nucleon decay does not have enough phase space to produce a charm or top quark. Consequently, this operator vanishes. In fact, this is the only operator in $C_6^{(Nm3)}$. Although this does not alter our results in the text, the usefulness of explicit operator construction is demonstrated, namely that symmetries and statistics may combine to forbid some operators, and sometimes an entire class of them.
\subsection{Operators for dinucleon to dilepton decays}

Here we list examples of operators from each class mediating dinucleon decays to dileptons (see table \ref{NN_class_table}). The superscript of an operator denotes the processes it mediates. For example, an operator ${\cal O}^{(nn,np,pp)}$ can mediate $nn \to \bar \nu \bar \nu^\prime$, $np \to \ell^+ \bar\nu$ and $pp \to \ell^+ \ell^{\prime +}$. Let us define the color tensors necessary to obtain a singlet from six fundamental representations of $SU(3)_c$. 
\begin{align}
    (T_s)_{\alpha \beta \gamma \delta \rho \sigma} \equiv{}&
\epsilon_{\rho \alpha \gamma}\epsilon_{\sigma \beta \delta} +
\epsilon_{\sigma \alpha \gamma}\epsilon_{\rho \beta \delta} + \epsilon_{\rho \beta \gamma}\epsilon_{\sigma \alpha \delta} +
\epsilon_{\sigma \beta \gamma}\epsilon_{\rho \alpha \delta} \label{Ts_eqn} \ , \\
(T_a)_{\alpha \beta \gamma \delta \rho \sigma} \equiv{}&
\epsilon_{\rho \alpha \beta}\epsilon_{\sigma \gamma \delta} +
\epsilon_{\sigma \alpha \beta}\epsilon_{\rho \gamma \delta} \label{Ta_eqn} \ . 
\end{align}
Let us also define threefold symmetric $SU(2)_{L,R}$ contraction tensors for six fundamental representations:
    \begin{multline}
    (I_{sss,L})_{ijkmpq} \equiv{}
\epsilon_{ik}\big (\epsilon_{jp}\epsilon_{mq}+\epsilon_{mp}\epsilon_{jp}\big )
+ 
\epsilon_{im}\big (\epsilon_{jp}\epsilon_{kq}+\epsilon_{kp}\epsilon_{jm}\big )
+ \\ +
\epsilon_{jk}\big (\epsilon_{ip}\epsilon_{mq}+\epsilon_{mp}\epsilon_{iq}\big )
+
\epsilon_{jm}\big (\epsilon_{ip}\epsilon_{kq}+\epsilon_{kp}\epsilon_{iq}\big ) \ , \label{Isss_eqn_L} 
    \end{multline}
    \begin{multline}
    (I_{sss,R})_{i'j'k'm'p'q'} \equiv{}
\epsilon_{i'k'}\big (\epsilon_{j'p'}\epsilon_{m'q'}+\epsilon_{m'p'}\epsilon_{j'p'}\big )
+ 
\epsilon_{i'm'}\big (\epsilon_{j'p'}\epsilon_{k'q'}+\epsilon_{k'p'}\epsilon_{j'm'}\big )
+ \\ +
\epsilon_{j'k'}\big (\epsilon_{i'p'}\epsilon_{m'q'}+\epsilon_{m'p'}\epsilon_{i'q'}\big )
+
\epsilon_{j'm'}\big (\epsilon_{i'p'}\epsilon_{k'q'}+\epsilon_{k'p'}\epsilon_{i'q'}\big )  \ . \label{Isss_eqn_R}
    \end{multline}
Let us now list one operator from each class:
\begin{multline}
       {\mathcal{O}}_1^{(pp,np,nn)} = {} \big[Q_{R}^{i^\prime \, \alpha  T} C Q_{R}^{j^\prime \beta} \big] \, \big[Q_{R}^{k^\prime \, \gamma  T} C Q_{R}^{m^\prime \delta} \big] \, \big[Q_{R}^{p^\prime \, \rho  T} C L_{R}^{q^\prime} \big] \, \big[Q_{R}^{r^\prime \, \sigma  T} C L_{R}^{s^\prime} \big] \times \\
       \times (T_a)_{\alpha \beta \gamma \delta \rho \sigma} \, \epsilon_{i'j'}  \, \epsilon_{k'm'} \, \epsilon_{p'q'} \, \epsilon_{r's'} \in C_1^{(N N^\prime)} \ ,
       \label{O1_NN_eqn}
\end{multline}
\begin{multline}
    {\mathcal{O}}_2^{(np)} = {} \big[Q_{R}^{i^\prime \, \alpha  T} C Q_{R}^{j^\prime \beta} \big] \, \big[Q_{R}^{k^\prime \, \gamma  T} C Q_{R}^{m^\prime \delta} \big] \, \big[Q_{R}^{p^\prime \, \rho T} C Q_{R}^{q^\prime \sigma} \big] \, \big[L_{L}^{i \, T} C L_{L}^{j} \big] \times \\
    \times \epsilon_{i j} \, (I_{sss,R})_{i'j'k'm'p'q'} \, (T_s)_{\alpha \beta \gamma \delta \rho \sigma} \in C_2^{(N N^\prime)} \ ,
    \label{O2_NN_eqn}
\end{multline}
\begin{multline}
    {\mathcal{O}}_3^{(pp,np,nn)} = {} \big[Q_{R}^{i^\prime \, \alpha  T} C Q_{R}^{j^\prime \beta} \big] \, \big[Q_{R}^{k^\prime \, \gamma  T} C Q_{R}^{m^\prime \delta} \big] \, \big[Q_{R}^{p^\prime \, \rho T} C L_{R}^{q^\prime} \big] \, \big[Q_{L}^{i \, \sigma T} C L_{L}^{j} \big] \times \\
    \times \epsilon_{i j} \, \epsilon_{i' j'} \, \epsilon_{k'm'} \,  \epsilon_{p'q'} \, (T_a)_{\alpha \beta \gamma \delta \rho \sigma} \in C_3^{(N N^\prime)} \ ,
    \label{O3_NN_eqn}
\end{multline}
\begin{multline}
    {\mathcal{O}}_4^{(pp,nn)} = {} \big[Q_{L}^{i \, \alpha  T} C Q_{L}^{j \beta} \big] \, \big[Q_{R}^{i^\prime \, \gamma  T} C Q_{R}^{j^\prime \delta} \big] \, \big[Q_{R}^{k^\prime \, \rho T} C Q_{R}^{m^\prime \sigma} \big] \, \big[L_{R}^{p^\prime \, T} C L_{R}^{q^\prime} \big] \times \\
    \times \epsilon_{i j} \, \epsilon_{i' j'} \, \big(\epsilon_{k'p'}\epsilon_{m'q'}+ \epsilon_{k'q'}\epsilon_{m'p'}\big) \, (T_a)_{\alpha \beta \gamma \delta \rho \sigma} \in C_4^{(N N^\prime)} \ ,
    \label{O4_NN_eqn}
\end{multline}
\begin{multline}
    {\mathcal{O}}_5^{(np)} = {} \big[Q_{R}^{i^\prime \, \alpha  T} C Q_{R}^{j^\prime \beta} \big] \, \big[Q_{R}^{k^\prime \, \gamma  T} C Q_{R}^{m^\prime \delta} \big] \, \big[Q_{L}^{i \, \rho T} C L_{L}^{j} \big] \, \big[Q_{L}^{k \, \sigma T} C L_{L}^{m} \big] \times \\
    \times \epsilon_{i' j'} \, \epsilon_{k' m'} \, \big(\epsilon_{i k}\epsilon_{j m}+ \epsilon_{i m}\epsilon_{j k}\big) \, (T_a)_{\alpha \beta \gamma \delta \rho \sigma} \in C_5^{(N N^\prime)} \ ,
    \label{O5_NN_eqn}
\end{multline}
\begin{multline}
    {\mathcal{O}}_6^{(np)} = {} \big[Q_{L}^{i \, \alpha  T} C Q_{L}^{j \beta} \big] \, \big[Q_{L}^{k \, \gamma  T} C Q_{L}^{m \delta} \big] \, \big[Q_{R}^{i' \, \rho T} C L_{R}^{j'} \big] \, \big[Q_{R}^{k' \, \sigma T} C L_{R}^{m'} \big] \times \\
    \times \epsilon_{i j} \, \epsilon_{k m} \, \big(\epsilon_{i' k'}\epsilon_{j' m'}+ \epsilon_{i' m'}\epsilon_{j' k'}\big) \, (T_a)_{\alpha \beta \gamma \delta \rho \sigma} \in C_6^{(N N^\prime)} \ ,
    \label{O6_NN_eqn}
\end{multline}
\begin{multline}
    {\mathcal{O}}_7^{(pp,np,nn)} = {} \big[Q_{L}^{i \, \alpha  T} C Q_{L}^{j \beta} \big] \, \big[Q_{R}^{i' \, \gamma  T} C Q_{R}^{j' \delta} \big] \, \big[Q_{L}^{k \, \rho T} C L_{L}^{m} \big] \, \big[Q_{R}^{k' \, \sigma T} C L_{R}^{m'} \big] \times \\
    \times \epsilon_{i j} \, \epsilon_{k m} \, \epsilon_{i' j'} \, \epsilon_{k' m'}  \, (T_a)_{\alpha \beta \gamma \delta \rho \sigma} \in C_7^{(N N^\prime)} \ ,
    \label{O7_NN_eqn}
\end{multline}
\begin{multline}
    {\mathcal{O}}_8^{(np)} = {} \big[Q_{L}^{i \, \alpha  T} C Q_{L}^{j \beta} \big] \, \big[Q_{L}^{k \, \gamma  T} C Q_{L}^{m \delta} \big] \, \big[Q_{L}^{p \, \rho T} C Q_{L}^{q \sigma} \big] \, \big[L_{R}^{i^\prime \, T} C L_{R}^{j^\prime} \big] \times \\
    \times \epsilon_{i'j'} \, (I_{sss,L})_{ijkmpq} \, (T_s)_{\alpha \beta \gamma \delta \rho \sigma} \in C_8^{(N N^\prime)} \ ,
    \label{O8_NN_eqn}
\end{multline}
\begin{multline}
    {\mathcal{O}}_9^{(pp,np,nn)} = {} \big[Q_{L}^{i \, \alpha  T} C Q_{L}^{j^ \beta} \big] \, \big[Q_{L}^{k \, \gamma  T} C Q_{L}^{m \delta} \big] \, \big[Q_{L}^{p \, \rho T} C L_{L}^{q} \big] \, \big[Q_{R}^{i' \, \sigma T} C L_{R}^{j'} \big] \times \\
    \times \epsilon_{i' j'} \, \epsilon_{i j} \, \epsilon_{k m} \,  \epsilon_{p q} \, (T_a)_{\alpha \beta \gamma \delta \rho \sigma} \in C_9^{(N N^\prime)} \ ,
    \label{O9_NN_eqn}
\end{multline}
\begin{multline}
    {\mathcal{O}}_{10}^{(pp,nn)} = {} \big[Q_{R}^{i' \, \alpha  T} C Q_{R}^{j' \beta} \big] \, \big[Q_{L}^{i \, \gamma  T} C Q_{L}^{j \delta} \big] \, \big[Q_{L}^{k \, \rho T} C Q_{L}^{m \sigma} \big] \, \big[L_{L}^{p \, T} C L_{L}^{q} \big] \times \\
    \times \epsilon_{i' j'} \, \epsilon_{i j} \, \big(\epsilon_{kp}\epsilon_{mq}+ \epsilon_{kq}\epsilon_{mp}\big) \, (T_a)_{\alpha \beta \gamma \delta \rho \sigma} \in C_{10}^{(N N^\prime)} \ ,
    \label{O10_NN_eqn}
\end{multline}
\begin{multline}
       {\mathcal{O}}_{11}^{(pp,np,nn)} = {} \big[Q_{L}^{i \, \alpha  T} C Q_{L}^{j \beta} \big] \, \big[Q_{L}^{k \, \gamma  T} C Q_{L}^{m \delta} \big] \, \big[Q_{L}^{p \, \rho  T} C L_{L}^{q} \big] \, \big[Q_{L}^{r \, \sigma  T} C L_{L}^{s} \big] \times \\
       \times \epsilon_{ij}  \, \epsilon_{km} \, \epsilon_{pq} \, \epsilon_{rs} \, (T_a)_{\alpha \beta \gamma \delta \rho \sigma}  \in C_{11}^{(N N^\prime)} \ .
       \label{O11_NN_eqn}
\end{multline}
Let us take one of the operators, ${\cal O}_6^{(np)}$, and illustrate the process mediated by it. Considering all first generation fermion fields for the moment, explicitly, 
\begin{multline}
    {\cal O}_6^{(np)}  = 4 \big(u_L^{\alpha T} \, C \, d_L^\beta \big) \, \big(u_L^{\gamma T} \, C \, d_L^\delta \big) \big\{\big(u_R^{\rho T} \, C \, \nu_R \big) \, \big(d_R^{\sigma T} \, C \, \ell_R \big)+ \\ + \big(d_R^{\rho T} \, C \, \ell_R \big) \, \big(u_R^{\sigma T} \, C \, \nu_R \big)\big\} \times (T_a)_{\alpha \beta \gamma \delta \rho \sigma} \ .
    \label{O6_explicit_eqn}
\end{multline}
Clearly, ${\cal O}_6^{(np)}$ can mediate $np \to \ell^+ \bar\nu$, thus the superscript. Interestingly, the left-right symmetry is evident in these operators. To illustrate, consider ${\cal O}_4^{(pp,nn)}$ and ${\cal O}_{10}^{(pp,nn)}$, which are related by the left-right symmetry; these mediate the same processes, and similarly for other pairs of operators. ${\cal O}_7^{(pp,np,nn)}$ is self left-right symmetric; therefore, we have an odd number of classes. Contrast this with $B-L$ violating $Nm3$ processes; for example, let us consider eqs. \eqref{O3_Nm3_eqn}, \eqref{O8_Nm3_eqn}, which mediate different processes. This is a result of the fact that $\Delta_R$ picks up a VEV, thereby breaking the left-right symmetry.  
\bibliographystyle{JHEP}
\bibliography{bvdlrs}

\end{document}